\documentclass[twocolumn]{aastex631}

\shorttitle{PDRs4all: JWST NIRSpec simulation}
\shortauthors{Canin et al.}

\usepackage[frozencache=true]{minted}

\begin{document}

\title{PDRs4all: NIRSpec simulation of integral field unit spectroscopy of the Orion Bar photodissociation region}

\author[0000-0002-7830-6363]{Am\'elie Canin}
\affiliation{Institut de Recherche en Astrophysique et Plan\'etologie (IRAP), University of Toulouse, France. \texttt{olivier.berne@irap.omp.eu}}

\author[0000-0002-1686-8395]{Olivier Bern\'e}
\affiliation{Institut de Recherche en Astrophysique et Plan\'etologie (IRAP), University of Toulouse, France. \texttt{olivier.berne@irap.omp.eu}}

\author{The PDRs4All ERS team}

\begin{abstract}
The James Webb Space Telescope (JWST) was launched on December 25 2021. This document presents a simulation of 
the Near Infrared Spectrograph (NIRSpec) observations of the Orion Bar which will be performed as part of the 
Early Release Sciences (ERS) program ``PDRs4all". The methodology to produce this data relies on the use 
of a direct forward model of the instrument applied to a synthetic scene of the Orion Bar, coupled to 
format matching in order to deliver data in JWST-pipeline data format. The resulting 3D cube for one order
is provided publicly, and is compatible with tools developed by the STScI (e.g. Cubeviz) and with the 
science enabling products developed by the PDRs4all team. This cube can be used as a template 
observation for proposers who would like to apply for NIRSpec observations of extended sources with JWST.

\end{abstract}


\section{Introduction} \label{sec:intro}
The James Webb Space Telescope (\citealt{gard2006}, JWST) is a space telescope developed jointly by NASA, European Space Agency (ESA) and Canadian Space Agency (CSA). This telescope, launched on December 25, 2021, has four main scientific focuses: ``The End of the Dark Ages: First Light and Reionization"; ``The Assembly of Galaxies"; ``The Birth of Stars and Protoplanetary Systems"; and ``Planetary Systems and the Origins of Life". Thirteen Early Release Science (ERS) programs have been selected to demonstrate the scientific capabilities of JWST, to provide public data to the community, to educate and inform the community regarding JWST's capabilities. This paper is part of this effort in the context of the ERS program “PDRs4all: Radiative feedback from massive stars”\footnote{\url{http://pdrs4all.org}} (ID1288) which focuses on observations of the Orion Nebula \citep{ber21}. This 40-hour program will make use of three instruments aboard JWST, and will dedicate about 12.71 hours to spectroscopy of the Orion Bar with the Near Infrared Spectrograph (NIRSpec, \citealt{bagn2007}). The NIRSpec instrument \citep{bagn2007} has 9 filters between 0.6$\mu$m and 5.27$\mu$m out of which we will use 3 in the PDRs4all ERS project.

In this paper, we present a method to simulate NIRSpec hyperspectral images of the Orion Bar as planned in this ERS project, in the JWST pipeline format. The goal of these simulations is to prepare the tools that post-process the pipeline output and to test the ecosystem of analysis tools developed for JWST data\footnote{\url{https://jwst-docs.stsci.edu/jwst-post-pipeline-data-analysis}}.
The paper is organised as follows: Section \ref{sec:instrument} gives an overview of the NIRSpec instrument. In Section \ref{sec:simulation}, we present how we simulate the NIRSpec image following \citet{guil2020}, and make it compatible with the NIRSpec output pipeline format.

\section{NIRSpec Spectroscopy of the Orion Bar} \label{sec:instrument}
The Near Infrared Spectrograph (NIRSpec, \citealt{bagn2007}) is one of the four JWST instruments. There are four observing modes of NIRSpec and we are specifically interested in the imaging spectroscopy with the Integral Field Unit (IFU)\footnote{For more information: \url{https://jwst-docs.stsci.edu/jwst-near-infrared-spectrograph}}. 

\begin{figure}
    \centering
    \includegraphics[height = 6cm]{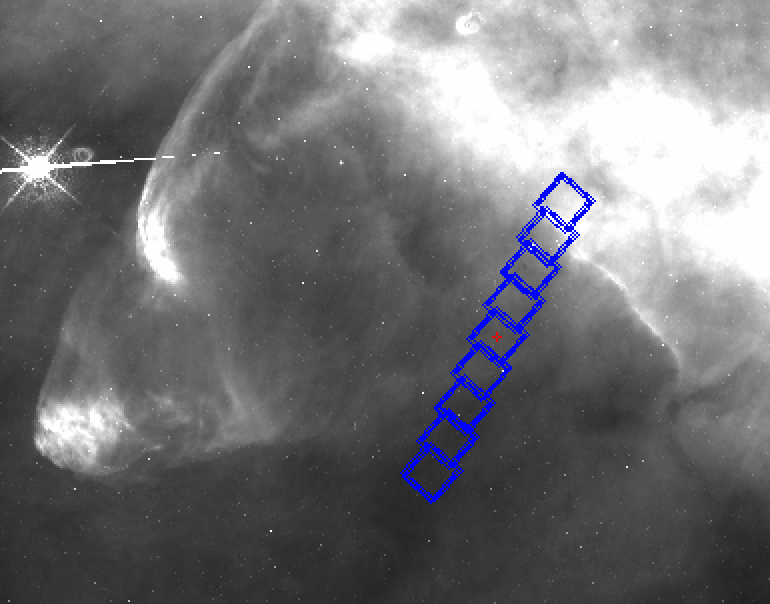}
    \caption{NIRSpec field of view for the PDRs4all project as specified in the ERS 1288 APT file, with HST-WFC3 \citep{kimble2008} with the F656N filter (6.56$\mu$m) image of the Orion star forming region in background. Blue regions: NIRSpec footprints corresponding to planned observations. Red cross: position of the target at the coordinates R.A.\ = 5:35:20.4749, dec.\ = -5:25:10.45.}
    \label{fig:nirspec_fov}
\end{figure}

The IFU mode has 9 disperser-filter combinations that span a total wavelength range of $0.6 \mu m$ to $ 5.3 \mu m$, and provide three levels of resolving power\footnote{More information on IFU mode: \url{https://jwst-docs.stsci.edu/jwst-near-infrared-spectrograph/nirspec-observing-modes/nirspec-ifu-spectroscopy}}. As part of the PDRs4all ERS project, 6 NIRSpec observations are planned with 3 disperser-filter combinations covering wavelengths between $0.97 \mu m$ and $ 5.27 \mu m$ with a nominal resolving power of 2,700. Each exposure will have one integration and each integration will consist of 5 groups with 4 dithers giving a total integration time of $257.68s$. The footprints of these NIRSpec observations as specified in the Astronomer's Proposal Tool (APT\footnote{ \url{https://jwst-docs.stsci.edu/jwst-astronomers-proposal-tool-overview}}) positioned over the Orion Bar are shown in Fig.\ \ref{fig:nirspec_fov}.

\begin{figure}
    \centering
    \includegraphics[height = 6cm]{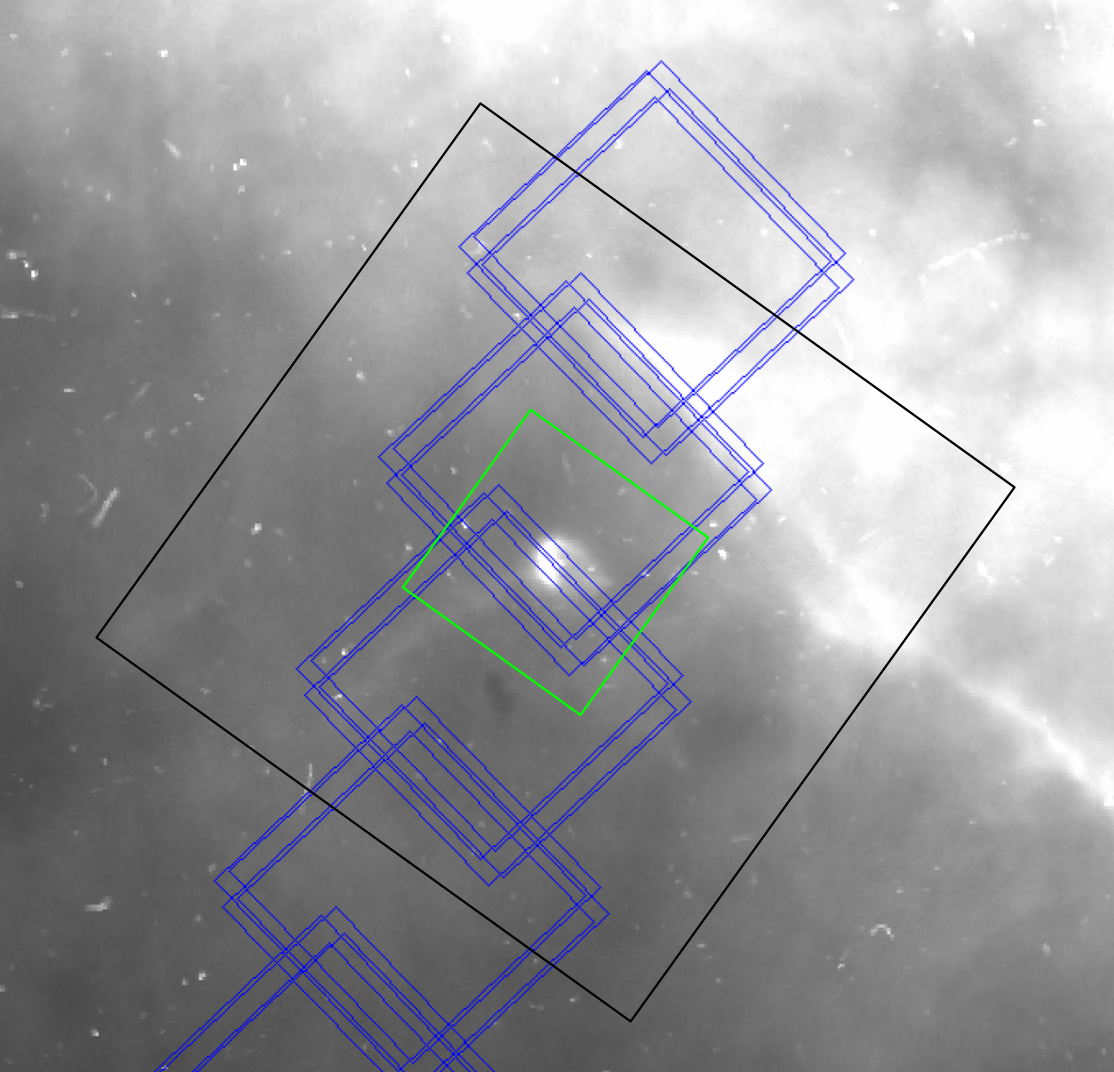}
    \caption{Zoom-in on the NIRSpec field of view from Fig.\ \ref{fig:nirspec_fov}. Blue: NIRSpec footprints corresponding to planned observations. Black: field of view of the simulation. Green: adopted field of view for the NIRSpec simulated cube.}
    \label{fig:zoom_fov}
\end{figure}

\section{Simulation} \label{sec:simulation}

\subsection{Motivation and strategy}

Creating NIRSpec simulations is useful to test JWST analysis tools developed by the PDRs4all team \citep{ber21} or by other teams including those of the STScI (e.g. Cubeviz, \citealt{cubeviz}). These simulations are also useful to obtain an idea of the quality (in terms of SNR) and richness of the data for a given integration time, ahead of observations. However, performing such simulations is challenging. There exists an instrument simulator \citep{piq10}, however simulating a full 3D NIRSpec cube (i.e. two spatial dimensions and one spectral dimension) with realistic spatial and spectral textures using this tool is very computationally intensive. 
As part of a project to develop algorithms to perform data-fusion between NIRSpec and NIRCam,  these authors have created a forward mathematical model of the NIRSpec instrument. They applied this forward model to a 3 dimensional input synthetic scene of the Orion Bar to create realistic NIRSpec simulations over the $1\mu m$ to $2.35 \mu m$ 
wavelength range. As part of the PDRs4all project, a larger wavelength range is expected to be observed ($0.97 \mu m$ to $5.2 \mu m$) with NIRSpec. In addition, \citet{guil2020} did not implement any tools to write out the cubes in the JWST pipeline format. In this paper, we present how we have extended the method of \citet{guil2020}.
To obtain a NIRSpec IFU simulated cube, we apply the direct model of 
\citet{guil2020} on the Orion Bar synthetic scene including the $0.97 \mu m$ to $5.2 \mu m$ range, and from this simulation we extract a cube with precisely the properties of the JWST-NIRSpec pipeline. We thus produce a realistic simulated IFU NIRSpec cube in the stage 3 format of the JWST-NIRSpec pipeline.

\subsection{Choice of region to be simulated}

Fig.\ \ref{fig:nirspec_fov} presents an overview of the footprints of the NIRSpec IFU observations planned in September 2022 on the Orion Bar as part of PDRs4all. They span a cut across the Orion Bar, performed with a mosaic strategy (see \citealt{ber21}). Fig.\ \ref{fig:zoom_fov} is a zoomed-in version of Fig \ref{fig:nirspec_fov} which includes additional information on the fields of view of the simulations. The black square shows the region over which we apply the direct model of \citet{guil2020} to the synthetic scene of the Orion Bar presented by the same authors.  The green square in Fig.~ \ref{fig:zoom_fov} corresponds to the field of view we adopted in the simulation of this paper.
It is a $3 \times 3''$ square (corresponding to NIRSpec IFU) centered on coordinates R.A.\ = 5:35:20.2570, dec.\ = -5:25:04.612. The orientation angle is $\frac{5}{9} \pi$ rad.
It overlaps with the planned mosaic, however we have centered it on one of the Proplyds, simply to help for coordinate calibration. We have only simulated one dither and one pointing. However, in principle, the method presented in this paper could be extended to simulate mosaics.

\subsection{Direct model of NIRSpec} \label{subsec:new_files}

\subsubsection{General principles of the model}

We follow and complement the formalism and notations of \citet{guil2020} to describe the forward mathematical model of NIRSpec that we will use.
We use a synthetic scene of the Orion Bar $\mathbf{C_i}$, which is a 3D cube sized $(12032 \times 300 \times 300)$, where 12032 is the number of spectral elements, and $300 \times 300$ the number of spatial elements. To compute this cube we define $\mathbf{X}$, which is a vectorized version 
of $\mathbf{C_i}$, sized $(12032 \times 90000)$, and  computed with the matrix product :  
\[ \mathbf{X} = \mathbf{H}\mathbf{A},
\]

where $\mathbf{H}$ is a matrix of elementary spectra sized $(12032 \times 4)$ and $\mathbf{A}$ is a matrix with the weight of the spectra spatially sized $(4 \times 90000)$.
$\bar \mathbf{Y}_{\text{h}}$, the hyperspectral NIRSpec image of size $(12032 \times 8649)$, is simulated using :

\[ \bar \mathbf{Y}_{\text{h}} = \mathbf{L}_{\text{h}}\mathcal{H}(\mathbf{X}) \mathbf{S} + \mathbf{N}, \]

where $\mathbf{L}_{\text{h}}$ is the NIRSpec throughput in a diagonal matrix, $\mathcal{H}(\cdot)$ is a spatial convolution with the JWST and NIRSpec point spread functions, which depends on wavelength and $\mathbf{S}$ is a downsampling operator corresponding to the spatial sampling of the NIRSpec instrument and $\mathbf{N}$ is the simulated noise (see details in \citealt{guil2020}).
$\bar \mathbf{Y}_{\text{h}}$ is then reshaped in a $(12032 \times 93 \times 93)$ 3D cube, $\mathbf{C_s}$. Finally, $\mathbf{C_s}$ is cropped to the spatial dimensions of NIRSpec simulation, i.e. $(12032 \times 30 \times 30)$
to obtain the final NIRSpec simulated cube $\mathbf{C_f}$. For each filter set, one cube  $\mathbf{C_f}$ is obtained, 
$\mathbf{C_f^{\rm G140H/F100LP}}$, $\mathbf{C_f^{\rm G235H/F170LP}}$ and $\mathbf{C_f^{\rm G395H/F290LP}}$.

\subsubsection{Contents of model matrices}

\paragraph{Matrices $\mathbf{A}$, $\mathbf{S}$, $\mathbf{N}$ } These matrices are computed in the same fashion as in \citet{guil2020}.

\paragraph{Matrix $\mathbf{H}$} This matrix contains the 4 elementary spectra which have been created as part of the PDRs4all project. The initial version of these 4 spectra was presented in \citet{guil2020}, however an updated version of these spectra is described in \citet{ber21}. Here we use the former version of $\mathbf{H}$. The total wavelength range is $0.7 \mu m - 5.2 \mu m$ for 12032 spectral points.

\paragraph{Matrix $\mathbf{L}_{\text{h}}$}  This matrix is a diagonal matrix where the diagonal corresponds to the throughputs of NIRSpec. \texttt{Pandeia} \citep{pont2016}, a Python package developped at STScI, is used. These package calculates the throughputs of the four JWST instruments. For NIRSpec, several inputs are needed: the mode, the disperser, the filter, the readout pattern, the number of integration and the number of groups. All this information is available in the ERS APT proposal (ID 1288, 'PDRs4all'). \texttt{Pandeia} also needs the wavelengths table on which to calculate the throughputs. In our case, the mode is \texttt{IFU}, the readout pattern is \texttt{nrsrapid}, there is 1 integration and 5 groups. The observations are planned with the following disperser-filter combinations: \texttt{G140H/F100LP}, \texttt{G235H/F170LP} and \texttt{G395H/F290LP}. So, 3 different curves are obtained depending on the dispersers/filters. These curves correspond to the diagonal of the matrix, so there are 3 matrices. The code to calculate these throughputs is presented in Listing \ref{lst:pce}. Fig.\ \ref{fig:pce_curves} presents the curves obtained for each disperser-filter combination and only for those used in the ERS program.

\begin{figure}
    \centering
    \includegraphics[height = 4 cm]{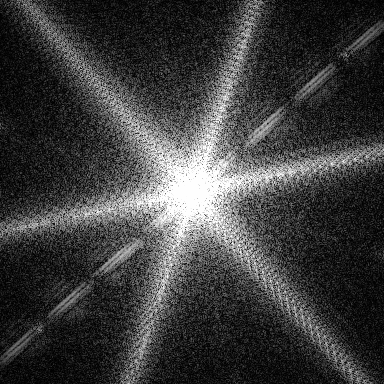}
    \includegraphics[height = 4 cm]{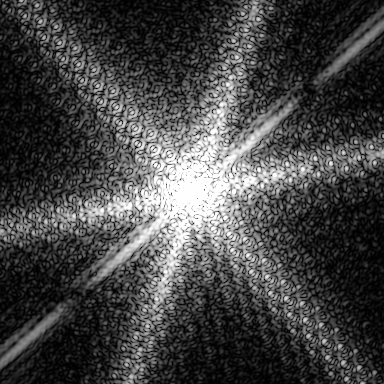}
    \caption{Real part of the NIRSpec fft PSF at $0.99 \mu m$ and $2.38 \mu m$.}
    \label{fig:fig_psf}
\end{figure}

\paragraph{Operator $\mathcal{H}(\cdot)$} The operator $\mathcal{H}(\cdot)$ is a convolution with point spread functions (PSFs), stored in a matrix we call $\mathbf{G}$. $\mathbf{G}$ has 4 dimensions and stores the fast Fourier transform (fft) of the NIRSpec PSFs. The two first dimensions are the spatial dimensions of the PSF. The third dimension is the spectral dimension and the last one is for the real part and the imaginary part of the NIRSpec PSF fft. 
\\First, we calculate the NIRSpec PSF with \texttt{Webbpsf} \citep{perr2014}, a Python package from the STScI. This package allows to calculate the PSF of NIRSpec for each spectral point. 
\\ Then, the fft of the PSFs cubes are calculated and saved in fits files. They are assembled to form one unique cube with all the wavelengths in the matrix $\mathbf{G}$. The code to calculate the PSF with \texttt{Webbpsf} is presented in Listing \ref{lst:psf}. Fig.\ \ref{fig:fig_psf} shows two examples of NIRSpec PSF obtained after the fft.

\subsubsection{Format matching} \label{subsec:format}

Here we consider the case of the \texttt{G140H/F100LP} filter.
We create a file in the stage 3 format of the pipeline, 
i.e. an \texttt{\_s3d} file in fits format. This file includes data 
and metadata. The data is comprised of several extensions. Extension 1 is the primary data, we use $\mathbf{C_f^{\rm G140H/F100LP}}$ cube interpolated on the spectral grid of the NIRSpec simulated cube for filter \texttt{G140H/F100LP} provided by the STScI\footnote{NIRSpec IFU data set at \url{https://www.stsci.edu/jwst/science-planning/proposal-planning-toolbox/simulated-data}}.  
Extension 2 is the error, here we use an error of 10\% of $\mathbf{C_f^{\rm G140H/F100LP}}$. 
Extension 3 is the data quality array codded on 
32 bits. For instance, 0 means that there are no problems with the pixel while a value of 513 ($2^9+2^0$) corresponds to a bad pixel (2$^0$) outside the science area of detector (2$^9$). Here we use a cube with dimensions of $\mathbf{C_f^{\rm G140H/F100LP}}$ with all elements set to 0 corresponding to good pixels only, since the simulation does not contain bad pixels or pixels with issues.

The metadata is composed of two fits headers, a primary header and a header for extension 1 of the data (primary image). We create these headers by making a copy of the header provided by STScI for NIRSpec IFU simulated observations.  The primary header contains the information related to the program such as the name of the mission, program, PI, etc. This header is common to all instruments, we replace the relevant information with that from the header created for our ERS program using the pipeline for NIRCam simulations \citep{canin2021}. In addition, information related to the target and the exposure is replaced manually using the information found in the APT proposal. In the image header, the information relative to the WCS parameters is replaced manually with the coordinates $(\texttt{CRVAL1}, \texttt{CRVAL2}) = (83.8343959,-5.4179437)$ at the reference point $(\texttt{CRPIX1}, \texttt{CRPIX2}) = (15,15)$. An extract of the image header is presented in Fig.\ \ref{fig:header}.

The file corresponding to this simulation can be downloaded at \citep{cube_nirspec}. The same approach allows to compute the NIRSpec IFU files for the other filters sets (i.e. \texttt{G235H/F170LP} and \texttt{G395H/F290LP}), provided one has the template format for these filters, which is not the case at the time we publish this document. 

\begin{figure}
    \centering
    \includegraphics[height = 4cm]{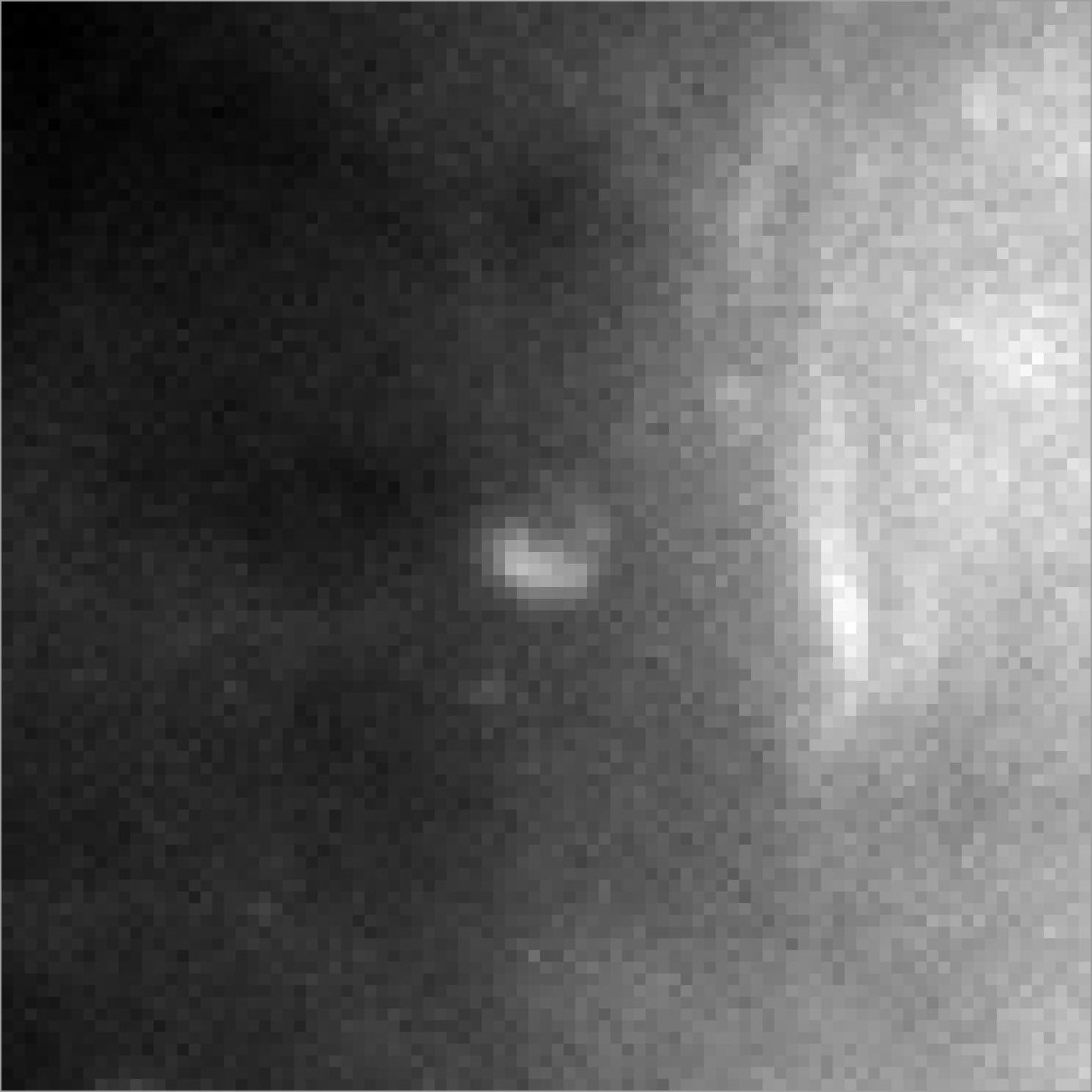}
    \includegraphics[height = 4cm]{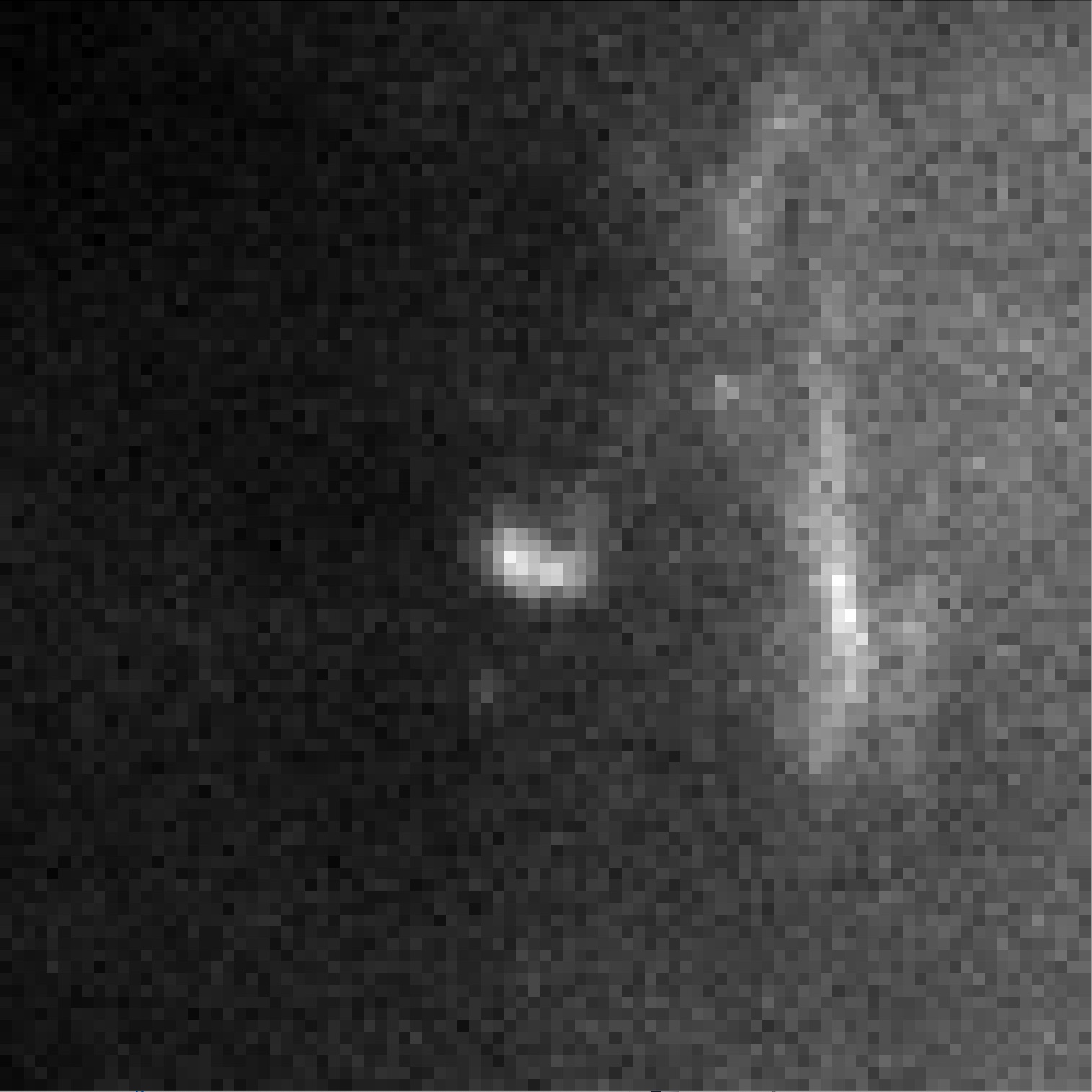}
    \caption{Images taken from the cube $\mathbf{C_s}$ at $0.97 \mu m$ and $2.87 \mu m$.}
    \label{fig:simu_scene}
\end{figure}

\section{Results}

\begin{figure}
    \centering
    \includegraphics[height = 10 cm]{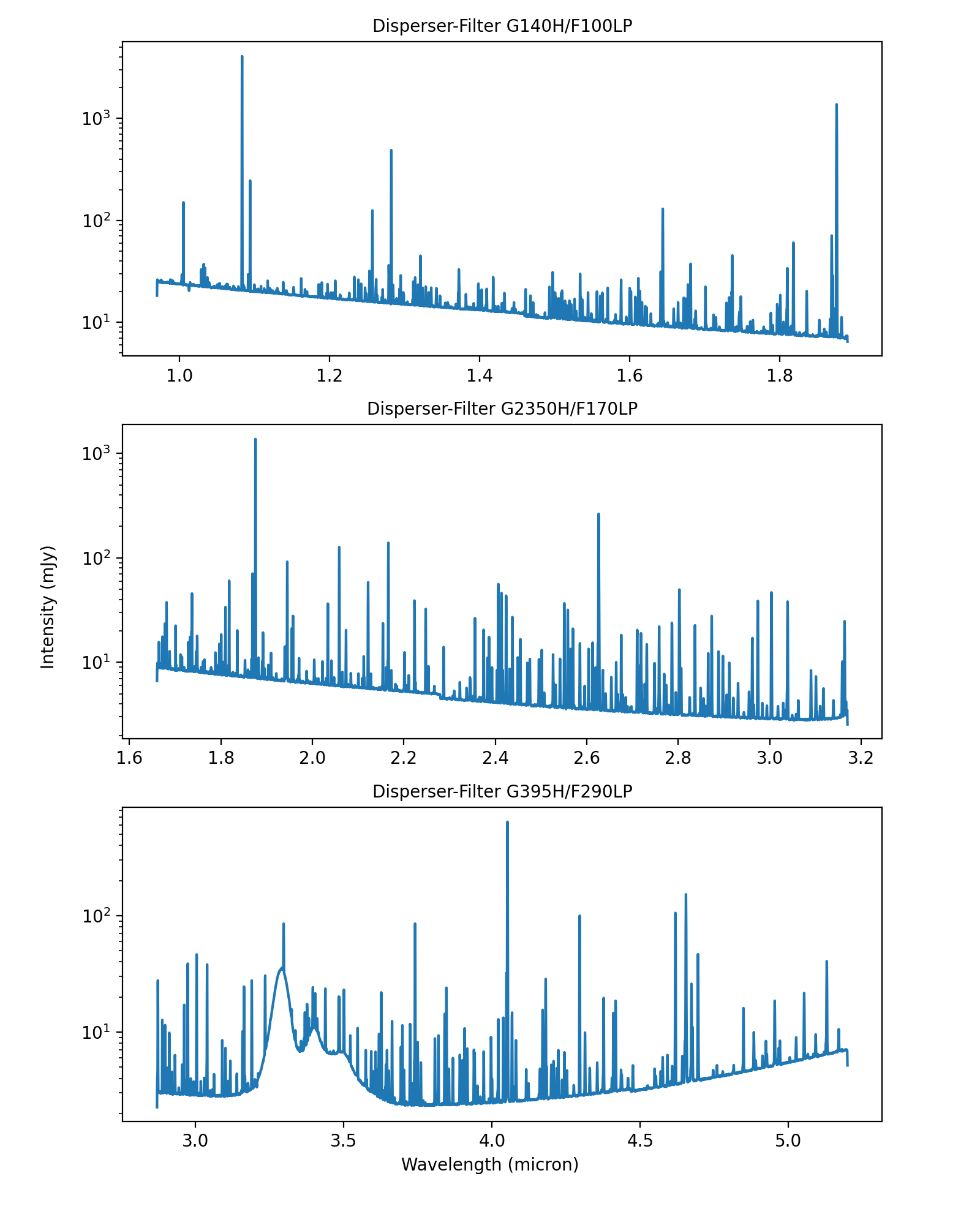}
    \caption{Spectra extracted from the $\mathbf{C_f}$ NIRSpec simulated cubes, corresponding to filters \texttt{G140H/F100LP}, \texttt{G235H/F170LP} and \texttt{G395H/F290LP}.}
    \label{fig:spectre}
\end{figure}

\begin{figure}
    \centering
    \includegraphics[height = 5 cm]{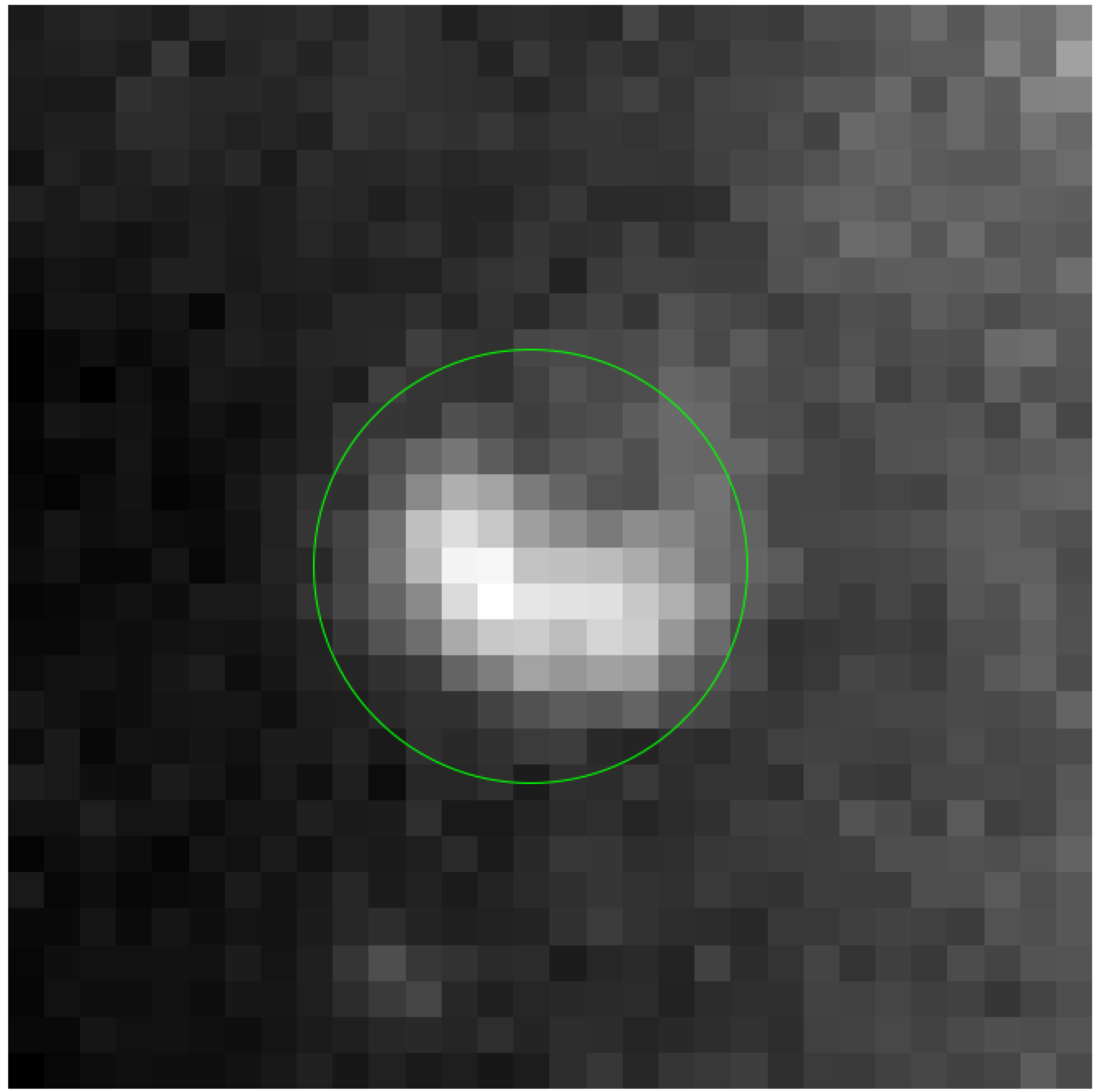}
    \caption{Image taken from the final NIRSpec simulated cube $\mathbf{C_f}$ at 2.12$\mu$m. The green region corresponds to the one on which the spectra in Fig.\ \ref{fig:spectre} were calculated.}
    \label{fig:img_crop}
\end{figure}

Fig.\ \ref{fig:simu_scene} presents the contents of the $\mathbf{C_s}$ cube for two wavelengths, $0.97 \mu m$ and $2.87 \mu m$. Fig.\ \ref{fig:spectre} presents the spectra of the three  $\mathbf{C_f}$ cubes extracted in the green circle region in Fig.\ \ref{fig:img_crop} which presents an image in of the final simulated cube $\mathbf{C_f}$, at $2.12\mu$m. This cube can be downloaded at \href{https://doi.org/10.5281/zenodo.5776707}{this link}.

\bibliography{sample631}{}

\begin{thebibliography}{}
\expandafter\ifx\csname natexlab\endcsname\relax\def\natexlab#1{#1}\fi
\providecommand{\url}[1]{\href{#1}{#1}}
\providecommand{\dodoi}[1]{doi:~\href{http://doi.org/#1}{\nolinkurl{#1}}}
\providecommand{\doeprint}[1]{\href{http://ascl.net/#1}{\nolinkurl{http://ascl.net/#1}}}
\providecommand{\doarXiv}[1]{\href{https://arxiv.org/abs/#1}{\nolinkurl{https://arxiv.org/abs/#1}}}

\bibitem[{{Bagnasco} {et~al.}(2007){Bagnasco}, {Kolm}, {Ferruit}, {Honnen},
  {Koehler}, {Lemke}, {Maschmann}, {Melf}, {Noyer}, {Rumler}, {Salvignol},
  {Strada}, \& {Te Plate}}]{bagn2007}
{Bagnasco}, G., {Kolm}, M., {Ferruit}, P., {et~al.} 2007, in Society of
  Photo-Optical Instrumentation Engineers (SPIE) Conference Series, Vol. 6692,
  Cryogenic Optical Systems and Instruments XII, ed. J.~B. {Heaney} \& L.~G.
  {Burriesci}, 66920M, \dodoi{10.1117/12.735602}

\bibitem[{Bern\'e {et~al.}(2022)Bern\'e, Habart, Peeters, Abergel, Bergin, \&
  Bernard-Salas}]{ber21}
Bern\'e, O., Habart, E., Peeters, E., {et~al.} 2022, PASP in prep

\bibitem[{Canin {et~al.}(2021)Canin, Berné, \& team}]{canin2021}
Canin, A., Berné, O., \& team, T. P.~E. 2021, PDRs4all: Simulation and data
  reduction of JWST NIRCam imaging of an extended bright source, the Orion Bar.
\newblock \doarXiv{2112.03106}

\bibitem[{Canin {et~al.}(2022)Canin, Berné, \& team}]{cube_nirspec}
---. 2022, NIRSpec IFU simulation of the Orion Bar with F100LP/G140H filter,
  Zenodo, \dodoi{10.5281/zenodo.5776707}

\bibitem[{{Gardner} {et~al.}(2006){Gardner}, {Mather}, {Clampin}, {Doyon},
  {Greenhouse}, {Hammel}, {Hutchings}, {Jakobsen}, {Lilly}, {Long}, {Lunine},
  {McCaughrean}, {Mountain}, {Nella}, {Rieke}, {Rieke}, {Rix}, {Smith},
  {Sonneborn}, {Stiavelli}, {Stockman}, {Windhorst}, \& {Wright}}]{gard2006}
{Gardner}, J.~P., {Mather}, J.~C., {Clampin}, M., {et~al.} 2006, \ssr, 123,
  485, \dodoi{10.1007/s11214-006-8315-7}

\bibitem[{{Guilloteau} {et~al.}(2020){Guilloteau}, {Oberlin}, {Bern{\'e}},
  {Habart}, \& {Dobigeon}}]{guil2020}
{Guilloteau}, C., {Oberlin}, T., {Bern{\'e}}, O., {Habart}, {\'E}., \&
  {Dobigeon}, N. 2020, \aj, 160, 28, \dodoi{10.3847/1538-3881/ab9301}

\bibitem[{Jones {et~al.}(2019)Jones, D'Avella, Robitaille, Earl, robelgeda,
  Kassin, Ferguson, Averbukh, Ogaz, Lim, Bradley, Hunkeler, \&
  Sipocz}]{cubeviz}
Jones, C., D'Avella, D., Robitaille, T., {et~al.} 2019, spacetelescope/cubeviz:
  v0.3 - Release v0.3, 0.3.0,  Zenodo, \dodoi{10.5281/zenodo.2616702}

\bibitem[{Kimble {et~al.}(2008)Kimble, MacKenty, O'Connell, \&
  Townsend}]{kimble2008}
Kimble, R.~A., MacKenty, J.~W., O'Connell, R.~W., \& Townsend, J.~A. 2008, in
  Space Telescopes and Instrumentation 2008: Optical, Infrared, and Millimeter,
  ed. J.~M.~O. Jr., M.~W.~M. de~Graauw, \& H.~A. MacEwen, Vol. 7010,
  International Society for Optics and Photonics (SPIE), 431 -- 442,
  \dodoi{10.1117/12.789581}

\bibitem[{{Perrin} {et~al.}(2014){Perrin}, {Sivaramakrishnan}, {Lajoie},
  {Elliott}, {Pueyo}, {Ravindranath}, \& {Albert}}]{perr2014}
{Perrin}, M.~D., {Sivaramakrishnan}, A., {Lajoie}, C.-P., {et~al.} 2014, in
  Society of Photo-Optical Instrumentation Engineers (SPIE) Conference Series,
  Vol. 9143, Space Telescopes and Instrumentation 2014: Optical, Infrared, and
  Millimeter Wave, ed. J.~{Oschmann}, Jacobus~M., M.~{Clampin}, G.~G. {Fazio},
  \& H.~A. {MacEwen}, 91433X, \dodoi{10.1117/12.2056689}

\bibitem[{Piquéras {et~al.}(2010)Piquéras, Legros, Pons, Legay, Ferruit,
  Dorner, Pécontal, Gnata, \& Mosner}]{piq10}
Piquéras, L., Legros, E., Pons, A., {et~al.} 2010, in Modeling, Systems
  Engineering, and Project Management for Astronomy IV, ed. G.~Z. Angeli \&
  P.~Dierickx, Vol. 7738, International Society for Optics and Photonics
  (SPIE), 407 -- 417, \dodoi{10.1117/12.856860}

\bibitem[{{Pontoppidan} {et~al.}(2016){Pontoppidan}, {Pickering}, {Laidler},
  {Gilbert}, {Sontag}, {Slocum}, {Sienkiewicz}, {Hanley}, {Earl}, {Pueyo},
  {Ravindranath}, {Karakla}, {Robberto}, {Noriega-Crespo}, \&
  {Barker}}]{pont2016}
{Pontoppidan}, K.~M., {Pickering}, T.~E., {Laidler}, V.~G., {et~al.} 2016, in
  Society of Photo-Optical Instrumentation Engineers (SPIE) Conference Series,
  Vol. 9910, Observatory Operations: Strategies, Processes, and Systems VI, ed.
  A.~B. {Peck}, R.~L. {Seaman}, \& C.~R. {Benn}, 991016,
  \dodoi{10.1117/12.2231768}

\end{thebibliography}
\bibliographystyle{aasjournal}

\appendix

\section{Listings}
\subsection{Command to calculate the throughputs} \label{lst:pce}
\begin{minted}[frame=single]{python}
obsmode = { 'instrument': 'nirspec',
            'mode': 'ifu',
            'disperser': 'g140h',
            'filter': 'f100lp'}

detector = {'readout_pattern': 'nrsrapid',
            'nint': 1,
            'ngroup': 5}

conf = {'instrument': obsmode, 'detector': detector}

i = pandeia.engine.instrument_factory.InstrumentFactory(config=conf)
pce = i.get_total_eff(tabwave)
\end{minted}

\subsection{Command to calculate the NIRSpec PSF} \label{lst:psf}
\begin{minted}[frame=single]{python}
nrs = webbpsf.NIRSpec()
nrs.image_mask = None # No MSA for IFU mode
wl = tabwave[start:end] # Cropped cube
cube = nrs.calc_datacube(wavelengths=wl, fov_pixels=fov_pixels)
\end{minted}

\section{Figures}
\begin{figure*}[h]
    \centering
    \includegraphics[height = 6 cm]{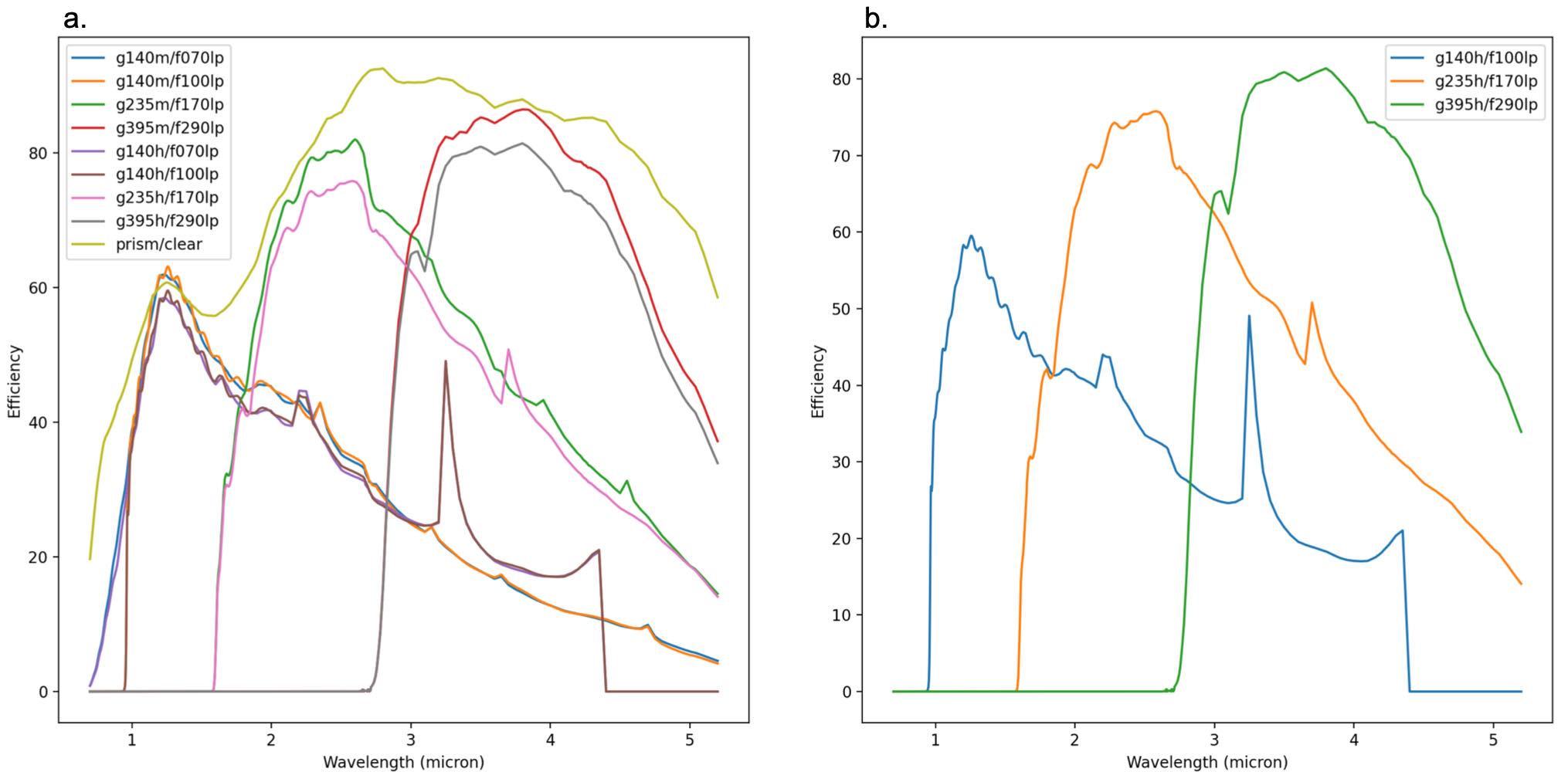}
    \caption{a. All the NIRSpec throughputs depending of the disperser-filter combination; b. The NIRSpec throughputs of the disperser-filters used in the ERS.}
    \label{fig:pce_curves}
\end{figure*}

\begin{figure*}[h]
\centering
\includegraphics[height = 12 cm]{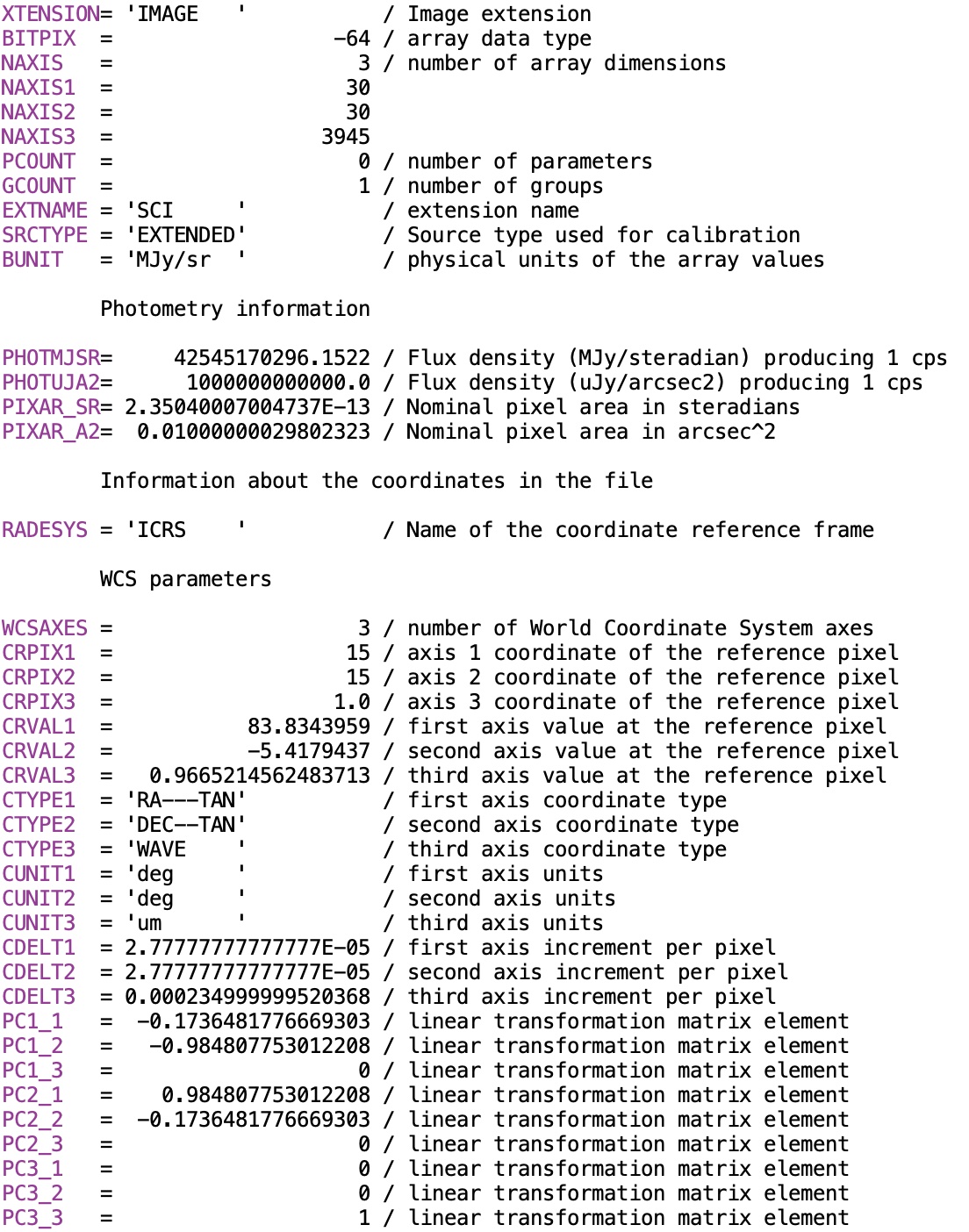}
\caption{Extract from the image header of the final NIRSpec simulated cube $\mathbf{C_f}$.}
\label{fig:header}
\end{figure*}

\end{document}